\documentclass[journal]{IEEEtran}
\usepackage{xcolor,soul,framed} 

\colorlet{shadecolor}{yellow}
\usepackage[pdftex]{graphicx}
\graphicspath{{../pdf/}{../jpeg/}}
\DeclareGraphicsExtensions{.pdf,.jpeg,.png}

\usepackage[cmex10]{amsmath}
\usepackage{bm}
\usepackage{amssymb}
\usepackage{array}
\usepackage{mdwmath}
\usepackage{mdwtab}
\usepackage{eqparbox}
\usepackage{url}
\usepackage{algorithm}  
\usepackage{algpseudocode}
\usepackage{tabularx}
\usepackage{tablefootnote}
\usepackage{threeparttable}
\usepackage{subfigure}
\usepackage{booktabs}
\usepackage{multicol}
\usepackage{multirow}
\usepackage{makecell}
\usepackage{setspace}

\begin{document}
\bstctlcite{IEEEexample:BSTcontrol}
    \title{Bridging Data-Driven and Model-Based Methods: A Learn-to-Optimize Architecture for Distributed Optimal Power Flow}
  \author{Yibo~Ding,~\IEEEmembership{Student Member,~IEEE,} 
      Zhao~Xu,~\IEEEmembership{Senior Member,~IEEE,} 
      Yuhong~Zhao,~\IEEEmembership{Student Member,~IEEE,} \\
      Jian~Zhao,~\IEEEmembership{Member,~IEEE,}
    Jiaqi~Ruan,~\IEEEmembership{Member,~IEEE}
    and~Zhaoyang~Dong,~\IEEEmembership{Fellow,~IEEE}}



\maketitle

\begin{abstract}
This letter proposes a learn-to-optimize (LTO) architecture for distributed optimal power flow (D-OPF) as the nexus between data-driven and model-based methods. By unfolding alternating direction method of multipliers (ADMM) into a deep neural network (NN) and embedding differentiable optimization layers, our architecture realizes near-instantaneous interpretable distributed decision-making. For mainstream relaxed formulations of D-OPF, the decisions from our architecture achieve comparable optimality with that of state-of-the-art solvers and excelled feasibility compared with existing data-driven approaches. Comparative case studies underpin the effectiveness of our architecture regarding the optimality and feasibility.
\end{abstract}


\IEEEpeerreviewmaketitle

\section{Introduction}
\IEEEPARstart{O}{ptimal} power flow (OPF) serves as the cornerstone of power systems operation. In practice, an interconnected power network or multiple mircogrids are often managed by distinct operators, making conventional centralized OPF (C-OPF) impractical due to managerial and privacy concerns. Then, D-OPF has emerged \cite{dall2013distributed}.

In this context, ADMM has been widely adopted in D-OPF \cite{erseghe2014distributed}. Based on explicit optimization models, ADMM yields satisfactory decision performance and preserves local data privacy, but requires iterative computing to converge. 
These drawbacks prohibit ADMM for near-instantaneous decision-making, which could be alternatively facilitated by data-driven methods (e.g. NN-based end-to-end (E2E) learning)\cite{dobbe2020toward}. However, D-OPF is a complex constrained optimization problem. Classical E2E learning acts as an uninterpretable black box that approximates the mapping from input features to final decisions and might offer physically infeasible solutions.

To enforce the feasibility of data-driven OPF solutions, the existing literature adopts mainly two strategies. The first leverages physics-informed NNs (PINNs) by exogenously penalizing constraint violations within the loss function \cite{lei2021data}, while the second introduces a post-processing stage in NN for remediation, such as projection optimization \cite{pan2021deepopf}. Intrinsically, these methods either use soft penalties to reduce but not eliminate constraints violations or execute post-hoc correction of potentially infeasible rough solutions. To actively and inherently generate feasible OPF solutions, embedding the underlying Karush-Kuhn-Tucker (KKT) conditions into the NN by constructing a differentiable optimization layer becomes promising \cite{agrawal2019differentiable}. Nevertheless, existing works that utilize differentiable optimization layers focus mainly on C-OPF \cite{jia2024optnet}, while D-OPF is never investigated.

One special challenge of data-driven D-OPF is the consensus on inter-regional tie-line power flow. To address this, deep unfolding (also called deep unrolling) technique could be adopted. By formulating the iterative primal and dual updates of ADMM as sequential NN layers, deep unfolding inherits the interpretablility of model-based methods and the near-instantaneous decision of data-driven methods\cite{noah2025distributed}. Existing deep unfolding implicitly embeds optimality conditions of the specific problem via intricate mathematical transformations of the closed-form solution (CFS), thereby fitting the expression of forward propagation (FP), such as the practice in state estimation using least-square residual formulation\cite{rout2025state}. However, CFSs rarely exist for OPF, severely hindering the application of deep unfolding in D-OPF.

To address these limitations, we propose a LTO architecture for D-OPF towards near-optimal, feasible, and fast distributed decision-making. Specifically, the ADMM-based deep unfolding framework is enabled for D-OPF as the first trial, where the embedded differentiable optimization layers perform primal updates, eliminating the reliance on deriving CFS. These layers rigorously integrate the KKT conditions of the convex-relaxed subproblems into the gradients through implicit differentiation. The decisions generated by our proposed architecture achieve optimality comparable with that of state-of-the-art solvers and exhibit significantly superior feasibility compared with conventional data-driven methods, thereby validating the interpretability of our architecture.

\section{Problem Formulation}\label{sec2}
Consider an interconnected power network containing a set of non-overlapping regions $\mathcal{R}=\{1,...,R\}$, the D-OPF aims to minimize total generation cost, subject to local physical constraints and inter-regional tie-line power flow consensus. In practice, direct current (DC)- or second-order conic programming (SOCP)-OPF are widely employed due to low computational burden and acceptable relaxation exactness \cite{low2014convex}. In this work, both formulations are examined. For each region $r\in \mathcal{R}$, the compact forms of DC- and SOCP-OPF are expressed in $\mathrm{(P1)}$ and $\mathrm{(P2)}$, respectively.
\allowdisplaybreaks
\begin{align}
& \mathrm{(P1)} \  \min \ f_r(\bm{x}_r) & \mathrm{(P2)} \ \min \ g_r(\bm{y}_r) \\
& \ \text{s.t.} \quad A_r \bm{x}_r = \mathbf{b}_r &  \ \text{s.t.}  \quad C_r \bm{y}_r = \mathbf{d}_r \\
& \quad \quad \ G_r \bm{x}_r \leq \mathbf{h}_r  & M_r \bm{y}_r \leq \mathbf{n}_r \\
& \quad \quad E_{r} \bm{x}_r = E_{s} \bm{x}_s & F_{r} \bm{y}_r = F_{s} \bm{y}_s \\
&  & \bm{y}_r \in \mathcal{K}_r  
\end{align} 
where $\bm{x}_r$ and $\bm{y}_r$ are the vectors collecting the decision variables of $\mathrm{(P1)}$ and $\mathrm{(P2)}$, respectively. The objectives $f_r(\cdot)$ and $g_r(\cdot)$ are typically quadratic cost functions of active power generation. $A_r$, $C_r$, $G_r$, and $M_r$ are constant matrices. $\mathbf{b}_r$ and $\mathbf{d}_r$ are constant vectors of equality constraints. $\mathbf{h}_r$ and $\mathbf{n}_r$ are constant vectors of physical limits. The third set of constraints (4) indicates the consensus on inter-regional tie-line power flow. $\bm{x}_s$ and $\bm{y}_s$ are the decision variables on the tie-line of the adjacent region $s$. $E_{r}$, $E_{s}$, $F_{r}$, and $F_{s}$ are the auxiliary mapping matrices. $\mathcal{K}_r$ is the second-order cone capturing the branch flow limits.

To solve (1) with ADMM, the primal and dual updates of region $r$ from iteration $k$ to $k+1$ are calculated as follows:
\begin{align} \label{eq.dc_primal}
    \textstyle \bm{x}_r^{k+1} = \arg \min \ f_r(\bm{x}_r) + \sum_{s} \frac{\rho}{2} || \bm{\varphi}_{rs} - \bm{\varphi}_{sr}^{k} + \bm{\lambda}_{rs}^{k} ||_2^2
\end{align}
\begin{equation} \label{eq.dc_dual}
    \bm{\lambda}_{rs}^{k+1} = \bm{\lambda}_{rs}^{k} + \rho(\bm{\varphi}_{rs}^{k+1} - \bm{\varphi}_{sr}^{k+1}), \quad \forall s
\end{equation}
\begin{align} \label{eq.socp_primal}
    \textstyle \bm{y}_r^{k+1} = \arg \min \ g_r(\bm{y}_r) + \sum_{s} \frac{\rho}{2} || \bm{\xi}_{rs} - \bm{\xi}_{sr}^{k} + \bm{u}_{rs}^{k} ||_2^2
\end{align}
\begin{equation} \label{eq.socp_dual}
    \bm{u}_{rs}^{k+1} = \bm{u}_{rs}^{k} + \rho(\bm{\xi}_{rs}^{k+1} - \bm{\xi}_{sr}^{k+1}), \quad \forall s
\end{equation}
where $\rho > 0$ is the penalty factor. $\bm{\varphi}_{rs}$ and $\bm{\xi}_{rs}$ are slack variables decoupling the consensus constraints for $\mathrm{(P1)}$ and $\mathrm{(P2)}$, respectively. $\bm{\lambda}_{rs}$ and $\bm{u}_{rs}$ are dual variables for inter-regional power flow consensus. The algorithm converges when the primal and dual residuals are sufficiently small.

\vspace{-5pt}
\section{The Proposed Architecture for D-OPF}\label{sec3}
\begin{figure}
    \centering
    \includegraphics[width=0.85\linewidth]{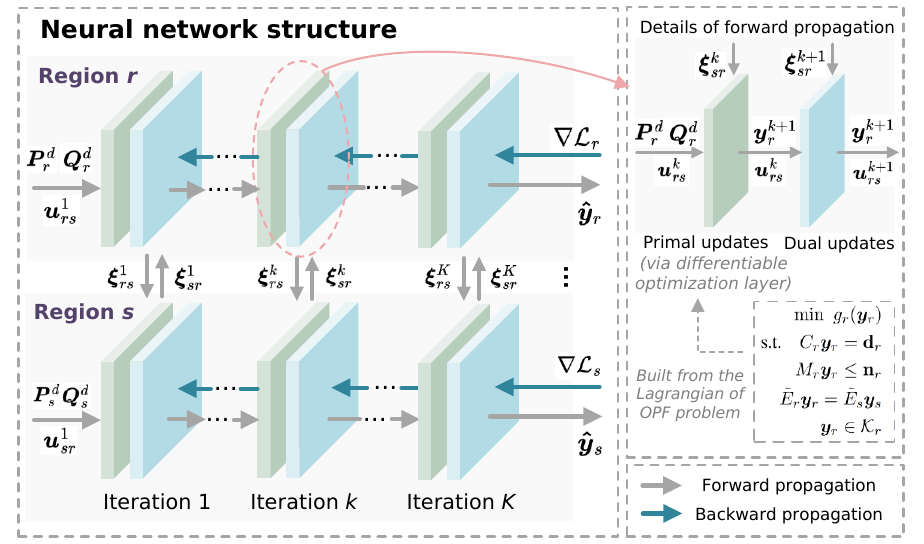}
    \caption{The proposed LTO architecture. The problem and variables are taken from SOCP-OPF formulation as an example.}
    \label{fig:Framework}
    \vspace{-12pt}
\end{figure}

Fig. \ref{fig:Framework} illustrates the proposed LTO architecture. 
Taking the $\mathrm{(P2)}$ as an example, the NN maps input active and reactive loads $[\bm{P}^d,\bm{Q}^d]$ to final decisions $\hat{\bm{y}}$ through iterative primal and dual updates \eqref{eq.socp_primal}-\eqref{eq.socp_dual}. Distinguished from existing methods, we formulate \eqref{eq.socp_primal} directly as a differentiable optimization layer \cite{agrawal2019differentiable}. By explicitly solving batched D-OPF problems during FP, this layer structurally guarantees high interpretability. Then, adjacent regions exchange only inter-regional variables and update dual variables via \eqref{eq.socp_dual}, where the other operational parameters are kept locally. Algorithm \ref{ag.learning} demonstrates the implementation of our architecture for D-OPF of SOCP formulation. Here, the penalty factor $\rho$ is set as learnable parameters of the NN. The mean square error (MSE) is employed as the loss function: $\mathcal{L}_r = \|\bm{y}_r - \hat{\bm{y}_r}\|^2/|\bm{y}_r|$.

During BP, we differentiate \eqref{eq.socp_primal} based on its KKT conditions to ensure feasibility. Here, $\mathrm{(P2)}$ with quadratic objective could be transformed into a standard SOCP with linear objective through epigraph reformulation. Then, the KKT conditions of $\mathrm{(P2)}$ are built into a homogeneous self-dual system (HSDS), a higher-dimensional set of linear equations. Then, by differentiating the HSDS, the KKT conditions are seamlessly embedded in the gradient in BP, and the physical constraints of D-OPF are enforced inherently by design, contributing to distinguished feasibility. Full mathematical detail could be refered to Ref. \cite{ding2025power}. Here, let $\Pi$ denote the mapping in FP from all the parameters $\bm{t}$ of $\mathrm{(P2)}$ to its optimal solution. The derivative of FP $\nabla_{\Pi}(\bm{t})$ could be decomposed into three composite mappings $\phi  \circ \nu  \circ \omega$, where $\circ$ is the function composition operator. Then, the derivative for BP is calculated as the adjoint $\nabla^\top _{\Pi}$ as follows.
\begin{equation}
    \nabla _{\Pi}(\bm{t})=\nabla_{\phi}(\nu(\omega)) \nabla_\nu(\omega(\bm{t})) \nabla_\omega(\bm{t})   \label{eq.forward_derivative}
\end{equation}
\begin{equation}
    \nabla^\top _{\Pi}(\bm{t})=\nabla^\top_\omega(\bm{t}) \nabla^\top_\nu(\omega(\bm{t})) \nabla^\top_{\phi}(\nu(\omega))   \label{eq.backward_derivative}
\end{equation}
where $\omega$ maps $\bm{t}$ to an auxiliary matrix that fits the mathematical form of HSDS. $\nu$ maps $\omega(\bm{t})$ to the solution of HSDS. $\phi$ recovers the optimal solution of $\mathrm{(P2)}$ from $\nu(\omega)$.


\begin{algorithm} 
    \caption{Implementation of the proposed architecture}
    \label{ag.learning}  
    \begin{algorithmic}[1]  
        \State \textbf{Dataset preparation:} Construct the full dataset $\mathcal{D}=\{\bm{P},\bm{Q},\bm{y}\}$ through probabilistic D-OPF, and separate into $\mathcal{D}^{\text{train}}_r$ and $\mathcal{D}^{\text{test}}_r$. Set the hyperparameters.
        \State \textbf{Network construction:} unfold the ADMM with $K$ iterations and build the differentiable optimization layers from the Lagrangian minimization problem (8);
        \State \textbf{Offline training:} (\textit{in parallel for all regions}) Input $\mathcal{D}^{\text{train}}_r$,
        \For{all the epochs}
            \State \textbf{Forward propagation:}
            \For{iteration $=1,...,K$}
                \State \textit{Primal updates}: optimize \eqref{eq.socp_primal} of the batched D-OPF
                \Statex \hspace{\algorithmicindent} $\qquad$ problems via differentiable optimization layers;
                \State \textit{Dual updates}: calculate \eqref{eq.socp_dual} after sharing the
                \Statex \hspace{\algorithmicindent} $\qquad$ inter-regional variables with adjacent regions;
            \EndFor
            \State \textbf{Backward propagation:}  Calculate the gradients of $\mathcal{L}_r$ with respect to all the learnable parameters;
        \EndFor
        \State \textbf{Online testing:} (\textit{for all regions}) Input $\mathcal{D}^{\text{test}}_r$, output $\hat{\bm{y}}_r$. Evaluate the decision performance.
    \end{algorithmic}
\end{algorithm}

\begin{table*}
\small
\centering
\caption{Local Decision Performances comparison of different methods.}
\label{tab.decision}
\begin{tabularx}{0.98\textwidth}{c|c|ccc|ccc|ccc|ccc}
\toprule
\specialrule{0em}{0.3pt}{0.5pt}
\toprule
\multirow{2}{*}{\centering \textbf{Model}} & \textbf{Method} & \multicolumn{3}{c|}{\textbf{M1}} & \multicolumn{3}{c|}{\textbf{M2}} & \multicolumn{3}{c|}{\textbf{M3}} & \multicolumn{3}{c}{\textbf{M4}} \\
\cmidrule(r){2-14}
& \textbf{Metric} &  \#1 & \#2 & \#3 
& \#1 & \#2 & \#3 
& \#1 & \#2 & \#3 
& \#1 & \#2 & \#3 \\

\midrule
 & Region 1 & 2.490\%  &	0.000  &	0.0\%  &	2.649\% & 0.000 & 0.0\% & 3.280\% & 	0.039 & 10.5\% &  2.036\% & 	0.249 & 	8.0\% \\
DC & Region 2 & 0.004\%  &	0.000 & 0.0\% & 0.027\% & 0.000 & 0.0\% &  0.446\% & 	0.012 & 22.0\% &  0.049\% & 	0.277 & 	25.0\%\\
 & Region 3 & 0.007\% & 	0.000 & 0.0\%  & 0.186\% & 0.000 & 0.0\% &  0.299\% & 	0.018  &	36.5\%  &  0.141\% & 	0.2009 & 	0.5\%\\
\midrule
 & Region 1 & 0.365\%  &	0.000  &	0.0\%  & 1.651\% & 0.000 & 0.0\% & 7.482\% & 	0.023 & 14.9\% & 2.248\% & 	0.223 & 	25.0\% \\
SOCP & Region 2 & 0.202\%  &	0.000 & 0.0\%  & 0.854\% & 0.000 & 0.0\% &  7.501\% & 	0.028 & 17.2\%  & 0.704\% & 	0.158 & 	27.5\%\\
 & Region 3 & 0.331\% & 	0.000 & 0.4\%  & 1.976\% & 0.000 & 0.0\% &  5.725\% & 	0.021  &	18.2\%  & 1.674\% & 	0.231 & 	34.2\%\\

\bottomrule
\specialrule{0em}{0.5pt}{0.3pt}
\bottomrule
\end{tabularx}
\end{table*}

\begin{table}
\footnotesize
\centering
\caption{Comparison of inter-region consensus error $\varepsilon$, training time $t^{tr}$ (s/epoch) and inference time $t^{inf}$ (ms/sample).}
\label{tab.time}
\begin{tabularx}{0.97\columnwidth}{c|ccc|ccc}
\toprule
\specialrule{0em}{0.3pt}{0.5pt}
\toprule
\multirow{2}{*}{\centering \textbf{Method}} & \multicolumn{3}{c|}{\centering DC-OPF}  & \multicolumn{3}{c}{\centering SOCP-OPF} \\
\cmidrule(r){2-7} 
 & $\varepsilon$ & $t^{tr}$ & $t^{inf}$ & $\varepsilon$ & $t^{tr}$ & $t^{inf}$ \\
\midrule
M1 & 1.74e-6 & 1105.2 & 57.727 & 6.16e-3 & 1541.2 & 79.166 \\
M2 & 1.25-4 & 89.77 & 7.1376 & 5.86-3 & 383.7 & 20.299 \\
M3 & 1.57e-4 & 129.2 & 0.4112 & 1.04e-2 & 421.5 & 1.6163 \\
M4 & 3.83e-5 & 32.27 & 0.3954 & 5.55e-3 & 35.07 & 0.6167 \\
Gurobi & 1.53e-6 & - & 2628.9 & 1.58-3 & - & 3580.2 \\

\bottomrule
\specialrule{0em}{0.5pt}{0.3pt}
\bottomrule
\end{tabularx}
\vspace{-12pt}
\end{table}

\vspace{-12pt}
\section{Cases Studies}\label{sec4}
Simulations are conducted on an IEEE-9 bus system managed by 3 operators. Given the parameters in Matpower files, results of ADMM-based probabilistic D-OPF solved by Gurobi serve as a learning benchmark, where a Guassian noise with 0.2 p.u. of standard deviation is added to power load. 10000 samples are partitioned into a training set and a testing set at an 8:2 ratio. $K$ and $\rho$ are set as 15 and 1, respectively. The other hyperparameters can be referred to Ref. \cite{ding2025power}. Compared to our approach (Method 1), we evaluate three ADMM-based deep unfolding baselines using standard multi-layer perceptron without differentiable optimization layers: Method 2 with post-processing remediation via projection optimization, and Methods 3 and 4 trained via physics-informed and pure MSE losses, respectively. To comprehensively evaluate the decision performance, our metrics include the mean relative error of generation cost that quantifies the optimality (Metric 1), the mean power flow equation error that evaluates the equality feasibility (Metric 2), and the inequality constraints violation rate (Metric 3) \cite{jia2024optnet}. In addition, the inconsistency of inter-regional power flow is considered as Metric 4. The hardware is i7-14700F CPU and NVIDIA GeForce RTX 4060Ti GPU on a PC.

Tables \ref{tab.decision} compares the local decision performance of these methods under DC- and SOCP-OPF formulations. Benefiting from the rigorous embedding of KKT conditions, Method 1 achieves the overall lowest optimality gap, yielding decisions that almost completely satisfy both equality and inequality constraints. Furthermore, as compared in Table \ref{tab.time}, it exhibits the lowest total inter-regional consensus error $\varepsilon$. Under the DC-OPF formulation, Method 2 achieves optimality comparable to Method 1 and higher inconsistency on inter-regional power flow. Conversely, in the SOCP-OPF formulation, although Method 2 shows a marginal improvement in inter-regional consistency, its optimality is notably inferior to that of Method 1. Furthermore, despite the increased structural complexity introduced by the post-processing stage, Method 2 generates decisions that almost fully satisfy both inequality and equality constraints. Compared to Method 4, Method 3 is trained with a physics-informed loss to enhance feasibility. However, relying solely on loss functions as soft penalties provides limited improvements in decision performance. Lacking a rigorous and inherent embedding of KKT conditions, the violation rates of both Methods 3 and 4 remain unacceptable.

Table \ref{tab.time} also compares the training time $t^{tr}$ (s/epoch) and inference time $t^{inf}$ (ms/sample) across all methods. Although the proposed method incurs a longer training time, this phase is executed entirely offline. More importantly, its inference time is significantly shorter than that of commercial solvers, satisfying real-time decision-making requirements. Our proposed architecture trades offline computational overhead for substantially improved optimality and feasibility, a critical advantage unattainable by conventional data-driven baselines.



\vspace{-5pt}
\section{Conclusion}\label{sec5}
A novel LTO architecture is proposed for D-OPF in this letter. Since the KKT conditions are seamlessly embedded into the gradients during BP, our architecture achieves near-optimal distributed decision-making with outperformed physical feasibility. The DC and SOCP relaxations could represent the majority cases of OPF in transmission network and distribution network, respectively. Moreover, our architecture is also applicable to other distributed optimization problems. 
Future work could move towards large-scale problems.

\ifCLASSOPTIONcaptionsoff
  \newpage
\fi

\bibliographystyle{IEEEtran}
\bibliography{IEEEabrv,Bibliography}
%

\vfill
\end{document}